\documentclass{PoS}

\title{$\Omega_c$ excited states with heavy-quark spin symmetry}

\ShortTitle{$\Omega_c$ excited states with heavy-quark spin symmetry}

\author{\speaker{Laura Tolos}\\
Institut f\"ur Theoretische Physik, University of Frankfurt, Max-von-Laue-Str. 1, 60438 Frankfurt am Main, Germany\\
Frankfurt Institute for Advanced Studies, University of Frankfurt, Ruth-Moufang-Str. 1,
60438 Frankfurt am Main, Germany\\
Institute of Space Sciences (ICE, CSIC), Campus UAB, Carrer de Can Magrans, 08193, Barcelona, Spain\\
Institut d'Estudis Espacials de Catalunya (IEEC), 08034 Barcelona, Spain \\
E-mail: \email{tolos@th.physik.uni-frankfurt.de}}

\author{Rafael Pavao\\
    Instituto~de~F\'{\i}sica~Corpuscular~(centro~mixto~CSIC-UV),
  Institutos~de~Investigaci\'on~de~Paterna, Aptdo.~22085,~46071,~Valencia,
  Spain\\}
  
  \author{Juan Nieves\\
    Instituto~de~F\'{\i}sica~Corpuscular~(centro~mixto~CSIC-UV),
  Institutos~de~Investigaci\'on~de~Paterna, Aptdo.~22085,~46071,~Valencia,
  Spain\\}

\abstract{We study the $C=1$, $S=-2$, $I=0$ sector, where five excited $\Omega_c$ states have been recently observed by the LHCb Collaboration. We start from a recently developed unitarized baryon-meson model that takes, as bare baryon-meson interaction, an extended Weinberg-Tomozawa kernel consistent with both chiral and heavy-quark spin symmetries. This ${\rm SU(6)} \times {\rm HQSS}$ scheme leads to a successful description of the observed lowest-lying odd parity charmed $\Lambda_c$(2595) and $\Lambda_c$(2625) states, and bottomed $\Lambda_b$(5912) and $\Lambda_b$(5920) resonances. Within this model, five odd-parity $\Omega_c$ states are dynamically generated, but with masses below 3 GeV, not allowing for an identification with the observed LHCb resonances. We revise this model and explore two different scenarios for the renormalization scheme, that is, using a modified common energy scale to perform the subtractions or utilizing a common ultraviolet cutoff to render finite the ultraviolet divergent loop functions in all channels. In both cases, we show that some (at least three) of the dynamically generated states can be identified with the experimental $\Omega_c$, while having odd parity and $J=1/2$ or $J=3/2$. Two of these states turn out to be part of the same ${\rm SU(6)} \times {\rm HQSS}$ multiplets as the charmed and bottomed $\Lambda$ baryons.}

\FullConference{XIII Quark Confinement and the Hadron Spectrum - Confinement2018\\
		31 July - 6 August 2018\\
		Maynooth University, Ireland}

\begin{document}

\section{Introduction}
\label{intro}

The LHCb Collaboration \cite{Aaij:2017nav} recently reported the existence of five excited $\Omega_c$ states, by means of analyzing the $\Xi_c^+ K^-$
decay in $pp$ collisions. With masses ranging between 3 and 3.1 GeV, four of these states have been also found by the Belle Collaboration \cite{Yelton:2017qxg}.   This discovery has triggered a large activity to determine their nature within quark models, QCD sum-rule schemes, quark-soliton models,
lattice QCD or molecular models. The final goal is to answer to the long-standing question whether these states can be explained within the quark model picture and/or these are molecules that are dynamically generated via hadron-hadron scattering.

Within  molecular models, there have been previous predictions on excited $\Omega_c$ states \cite{JimenezTejero:2009vq,Hofmann:2005sw,GarciaRecio:2008dp, Romanets:2012hm}. In view of the new discoveries, in Ref.~\cite{Montana:2017kjw} two $\Omega_c$ resonant states  at 3050 MeV and 3090 MeV were generated with spin-parity $J^P=1/2^-$, reproducing  two of the experimental states. Also, within an extended local hidden gauge
approach that incorporates  low-lying $1/2^+$ and $3/2^+$ baryons together with
pseudoscalar and vector mesons \cite{Debastiani:2017ewu},  two  $J^P=1/2^-$ $\Omega_c$ states  and one $J^P=3/2^-$ $\Omega_c^*$  were observed, the first two in good agreement with the outcome of \cite{Montana:2017kjw}.

With the aim of incorporating explicitly heavy-quark spin symmetry (HQSS), a proper QCD symmetry that
appears when the quark masses become larger
than the typical confinement scale, a scheme has been developed in
Refs.~\cite{GarciaRecio:2008dp,Gamermann:2010zz,Romanets:2012hm,GarciaRecio:2012db,Garcia-Recio:2013gaa,Tolos:2013gta,Garcia-Recio:2015jsa}
that implements a consistent ${\rm SU(6)}_{\rm lsf} \times {\rm HQSS}$ extension of the Weinberg-Tomozawa (WT) 
interaction, where ``lsf'' refers to light quark-spin-flavor symmetry. In fact, Refs.~\cite{GarciaRecio:2008dp, Romanets:2012hm} are the first
baryon-meson molecular analyses, fully consistent with HQSS, of the odd-parity $\Lambda_c(2595)$ [$J=1/2$] and
$\Lambda_c(2625)$ [$J=3/2$] resonances. The same model also generates dynamically the $\Lambda_b(5912)$ and $\Lambda_b(5920)$ narrow resonances, 
found by LHCb \cite{Aaij:2012da}, which are HQSS partners \cite{GarciaRecio:2012db,Garcia-Recio:2015jsa}. It turns out that the $\Lambda_b(5920)$ resonance
is the bottom counterpart of the $\Lambda_c(2625)$, whereas the $\Lambda_b(5912)$ is the bottomed version of the second charmed
state that is part of the two-pole structure of the $\Lambda_c(2595)$ ~\cite{GarciaRecio:2012db,Garcia-Recio:2015jsa}.

Within this scheme consistent with both chiral symmetry and HQSS,  five $\Omega_c$ states were found in Ref.~\cite{Romanets:2012hm}, three $J=1/2$ and the two $J=3/2$ bound states. These five odd-parity
$\Omega_c, \Omega^*_c$ states, stemming from the most attractive ${\rm
  SU(6)}_{\rm lsf}\times$ HQSS representations, have 
masses below 3 GeV, and cannot be easily identified with the LHCb
resonances. Predicted masses, however,  depend  not only on the baryon-meson interactions, but also
on the adopted renormalization scheme (RS). 

In this paper we review the RS used in  Ref.~\cite{Romanets:2012hm} and its impact in the generation of the five $\Omega_c^{(*)}$ states. We show how the pole positions can be moved up in energy by implementing a different RS, so as to make then  the identification of at least three states with the experimental $\Omega_c^{(*)}$ states feasible \cite{Nieves:2017jjx}.

\section{Unitarized baryon-meson model with ${\rm SU(6)} \times {\rm HQSS}$ Weinberg-Tomozawa kernel}
\label{unitarized}

We consider the charm $C=1$, strangeness $S=-2$ and isospin $I=0$ sector, where the $\Omega_c^{(*)}$ excited states are located, as we  revisited 
Ref.~\cite{Romanets:2012hm}. Starting from the baryon-meson pair of pseudoscalar and vector mesons as well as the low-lying $1/2^+$ and $3/2^+$ baryons in the  $C=1, S=-2, I=0$ sector, the $s$-wave  ${\rm SU(6)}_{\rm lsf}$ $\times$ HQSS WT  kernel is used as potential $V^J$ (for a given total angular momentum $J$) for the Bethe-Salpeter equation (BSE). This leads to a $T$-matrix 
\begin{equation}
\label{eq:LS}
T^J(s)=\frac{1}{1-V^J(s) G^J(s)} V^J(s),
\end{equation}
where $G^J(s)$ is a diagonal matrix that contains the loop functions for the different baryon-meson pairs. The  loop function is
logarithmically ultraviolet (UV) divergent and needs to be
renormalized. This can be done by one-subtraction 
\begin{equation}
G_i(s)=\overline{G}_i(s)+G_i(s_{i+}) ,
\label{eq:div}
\end{equation}
with the finite part of the loop function, $\overline{G}_i(s) $ \cite{Nieves:2001wt}. The divergent contribution, $G_i(s_{i+})$, needs to be renormalized.  Two different renormalization schemes are widely used. On the one hand,
the one subtraction at certain scale, $\sqrt{s}=\mu$, such that
\begin{eqnarray}
&&G_i(\sqrt{s}=\mu) = 0 ,  \\
&&G_i^\mu(s_{i+})  = - \overline{G}_i(\mu^2) ,  \\
&&G_i^\mu(s) =\overline{G}_i(s) - \overline{G}_i(\mu^2).
\label{eq:relation}
\end{eqnarray}
On the other hand, we make finite the UV divergent part of
the loop function using a sharp-cutoff regulator $\Lambda$ in momentum space, so  we have
\begin{equation}
G^{\Lambda}_i(s) =\overline{G}_i(s) + G_i^{\Lambda}(s_{i+}). \label{eq:uvcut2}
\end{equation}
Note that if one uses channel-dependent cutoffs, the one-subtraction RS, $\mu-RS$,  is recovered by
choosing in each channel $\Lambda_i$ such that
\begin{equation}
G^{\Lambda_i}_i(s_{i+})= -\overline{G}_i(\mu^2) .
\label{eq:subtraction}
\end{equation}
However, if one uses a common UV cutoff in a given $CSI$ sector, both RSs are independent and will lead to
different results.

The different dynamically-generated excited $\Omega_c^{(*)}$ are obtained as poles of the
scattering amplitudes in each $J$ sector for $(C=1,S=-2, I= 0)$.  We look at both
the first and second Riemann sheets (FRS and SRS) of the variable $\sqrt{s}$, with the following prescription: the poles of the
FRS that appear on the real axis below threshold are interpreted as bound states, whereas the poles 
on the SRS below the real axis and above threshold are identified with resonances. The mass and the
width can be found from the position of the pole on the complex energy plane:
\begin{equation}
T_{ij}(s) \simeq \frac{g_i g_j}{\sqrt{s}-\sqrt{s_R}},
\end{equation}
where the quantity $\sqrt{s_R}=M_R - \rm{i}\, \Gamma_R/2$ provides the mass ($M_R$) and the width
($\Gamma_R$) of the state, and $g_i$ is the complex coupling of the resonance to the channel $i$.

\section{Excited $\Omega_c^{(*)}$ states}
\label{excite}

%\begin{figure}
%     \includegraphics[width=.6\textwidth]{figure.eps}
%     \caption{This is the caption of the figure.}
%     \label{fig1}
 %    \end{figure}
     
%     \begin{table}
 %    \begin{tabular}{...}
 %    ...
 %    \end{tabular}
 %    \caption{This is the caption of the table.}
 %    \label{tab1}
 %    \end{table}

The LHCb Collaboration studied the $\Xi_c^+ K^-$ spectrum using $pp$
collisions and five new narrow excited $\Omega_c^0$ states were
identified: the $\Omega_c^0(3000)$, $\Omega_c^0(3050)$,
$\Omega_c^0(3066)$, $\Omega_c^0(3090)$ and the $\Omega_c^0(3119)$, the
last three also observed in the $\Xi_c^{'+} K^-$ decay. Moreover, a sixth
broad structure with 3188 MeV was found in the $\Xi_c^+ K^-$ spectrum. 

\subsection{One-subtraction renormalization}

\begin{table}[t]
  \centering
  \caption{ $\Omega_c$ an $\Omega_c^*$ excited states as reported in
    Ref.~\cite{Romanets:2012hm}. We label them from {\bf a} to {\bf e}, according to their energy position (taken from \cite{Nieves:2017jjx}).}
  \label{tab:table1}
  \begin{tabular}{c|c|c|c|}
      Name & $M_R$ (MeV) & $\Gamma_R$ (MeV) & $J$  \\
    \hline
    {\bf a} & 2810.9 & 0 & 1/2  \\
    \hline
    {\bf b} & 2814.3 & 0 & 3/2  \\
    \hline
    {\bf c} & 2884.5 & 0 & 1/2  \\
    \hline
    {\bf d} & 2941.6 & 0 & 1/2  \\
    \hline
   {\bf e} & 2980.0 & 0 & 3/2  \\
  \end{tabular}
\end{table}

\begin{table}[t]
  \centering
  \caption{ $\Omega_c$ and $\Omega_c^*$ excited states obtained using $\alpha=1.16$ (taken from \cite{Nieves:2017jjx}).}
  \label{tab:table2}
  \begin{tabular}{c|c|c|c||c|c}
      Name & $M_R$ (MeV) & $\Gamma_R$ (MeV) & $J$ & $M_R^{exp}$ & $\Gamma_R^{exp}$ \\
    \hline
    {\bf a} & 2922.2 & 0 & 1/2 & --- & --- \\
    \hline
    {\bf b} & 2928.1 & 0 & 3/2 & --- & --- \\
    \hline
   {\bf c} & 2941.3 & 0 & 1/2 & --- & --- \\
    \hline
    {\bf d}& 2999.9 & 0.06 & 1/2 & 3000.4 & 4.5 \\
    \hline
    {\bf e} & 3036.3 & 0 & 3/2 &  3050.2 & 0.8 \\
  \end{tabular}
\end{table}

As previously mentioned, in Ref.~\cite{Romanets:2012hm} five excited
 $\Omega_c$ states with $J=1/2^-$ and $J=3/2^-$  were predicted, with masses
below 3 GeV (Table~\ref{tab:table1}). The position in mass makes it difficult to identify any of them with 
 the LHCb resonances.

These states were dynamically generated by solving a coupled-channel BSE using a ${\rm SU(6)}_{\rm lsf}\times$ HQSS-extended
WT interaction as a kernel, whereas the baryon-meson loops were renormalized with one-substraction at the scale $\mu = \sqrt{\alpha \left(m_{th}^2+M_{th}^2 \right)}$, with $\alpha=1$, and $m_{th}$ and $M_{th}$ the masses of the meson and baryon of
the channel with the lowest threshold in the given $CSI$ sector~\cite{Hofmann:2005sw}. However, it is possible to permit some freedom
and slightly modify the choice of the subtraction point by changing  $\alpha$.  Allowing for just moderately changes, we find that for $\alpha=1.16$ the two
last states ({\bf d} and {\bf e} in Table~\ref{tab:table1}) are now located near the experimental
$\Omega_c(3000)$ and $\Omega_c(3050)$ (see Table \ref{tab:table2}).  The state with mass 2999.9 MeV is mainly generated by  $\Xi_c^{'+} \bar K$ , whereas the state at 3036.3 MeV has a dominant $\Xi_c^* \bar K$  component  that can be reconciled with the experimental decay $\Xi_c^+ K^-$ if we allow for  $\Xi_c^* \bar K \to
\Xi_c K$ $d-$wave transition \cite{Nieves:2017jjx}.

Given our results, we need to explore a different RS to evaluate the impact of the renormalization procedure in the predictions of  $\Omega_c^{(*)}$ in a controlled manner. Thus, we use the relation between the subtraction constants and the cutoff scheme, and employ a common UV
cutoff for all baryon-meson loops within reasonable limits. In this manner, we avoid any fictitious reduction of any
baryon-meson channel with the use of a small  cutoff value and we prevent an arbitrary variation of the subtraction constants.

\subsection{Common cutoff regularization}

\begin{table}[h]
  \centering
  \caption{ $\Omega_c$ and $\Omega_c^*$ excited states calculated using the
    subtraction constants from a cutoff of $\Lambda=1090$
    MeV (taken from \cite{Nieves:2017jjx}). }
  \label{tab:table3}
  \begin{tabular}{c|c|c|c||c|c}
      Name & $M_R$ (MeV) & $\Gamma_R$ (MeV) & $J$ & $M_R^{exp}$ & $\Gamma_R^{exp}$ \\
    \hline
    {\bf a} & 2963.95 & 0.0 & 1/2 & --- & --- \\
    \hline
    {\bf c} & 2994.26 & 1.85 & 1/2 & 3000.4 & 4.5 \\
    \hline
    {\bf b} & 3048.7 & 0.0 & 3/2 & 3050.2  &  0.8 \\
    \hline
    {\bf d} & 3116.81 & 3.72 & 1/2 & 3119.1/ 3090.2  & 1.1/ 8.7   \\
    \hline
    {\bf e} & 3155.37 & 0.17 & 3/2 &  --- & --- \\
  \end{tabular}
\end{table}

In order to identify our five dynamically generated
$\Omega_c^{(*)}$ of Table~\ref{tab:table1} using the new
subtraction constants, we first have to determine how their masses and
widths  change as we adiabatically vary the values of the subtraction constants. This is done by
\begin{equation}
G_i(s) = \overline{G}_i(s)-(1-x) \overline{G}_i(\mu^2)+x G_i^{\Lambda}(s_{i+}),
\end{equation}
where $x$ is a parameter that changes adiabatically from $0$ to $1$, and
$\mu^2=(m_{th}^2+M_{th}^2)$.  In this way, we can follow the original $\Omega_c^{(*)}$ in the
complex energy plane as we modified our prescription to use a common cutoff for the computation
of the subtraction constants.

In Table~\ref{tab:table3} we show our results for $\Omega_c^{(*)}$  for a cutoff of $\Lambda=1090$ MeV. We find that three poles (those named {\bf c},
{\bf b} and {\bf d}) can be identified with the three experimental
states at 3000 MeV, 3050 MeV, and 3119 or 3090 MeV. This identification is due to the closeness in energy to the experimental
states and because of the dominant contribution to their dynamical generation by  the experimental
$\Xi_c \bar K$ and $\Xi_c^{'} \bar K$ channels.

\begin{figure*}[ht!]
  \includegraphics[scale = 0.4]{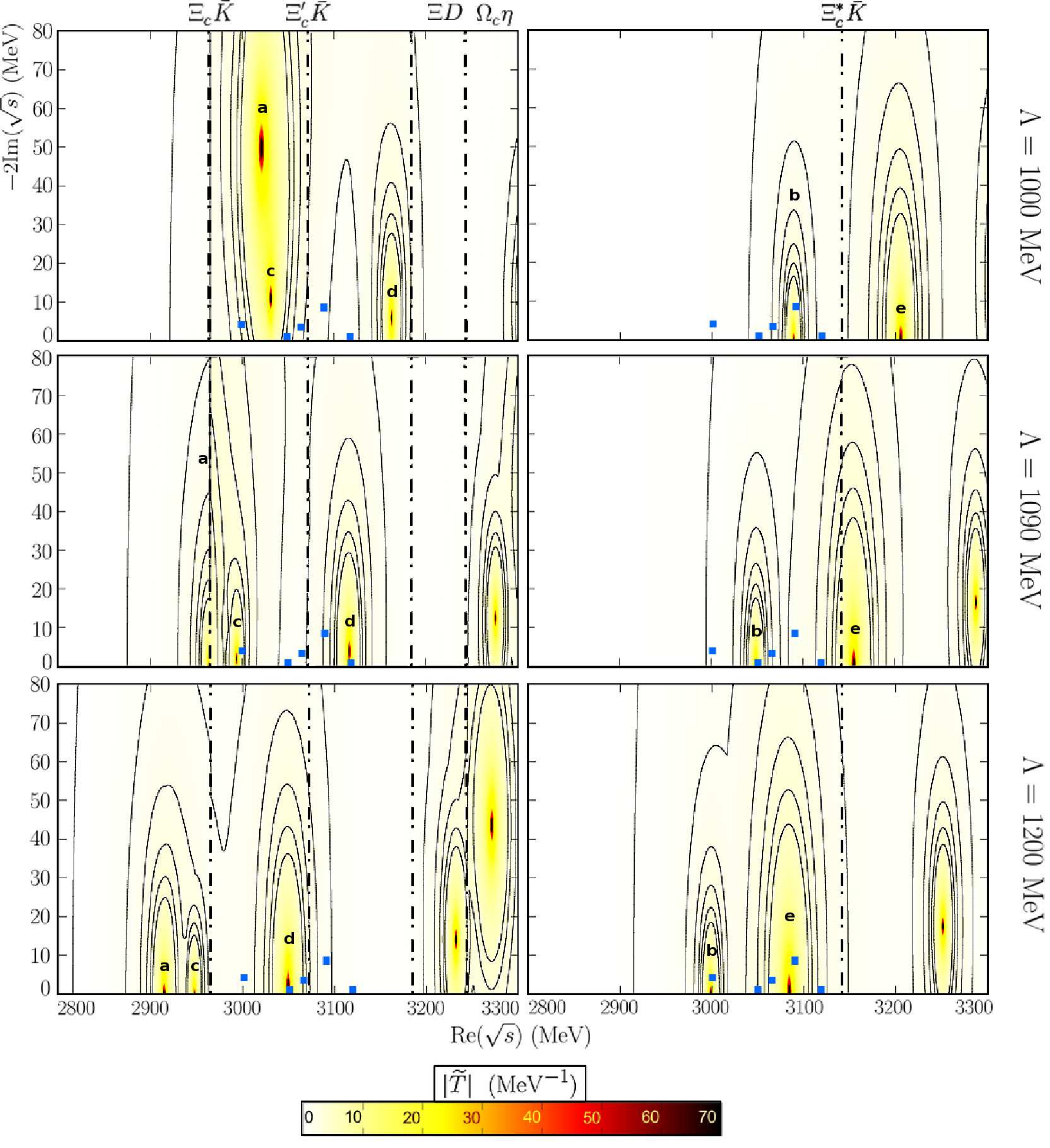}
%  \captionsetup{justification=raggedright}
  \caption{$\Omega_c$ and $\Omega_c^*$ excited states for different UV
    cutoffs.  The blue squares indicate the experimental
    results, whereas dashed-dotted lines show the closest baryon-meson
    thresholds. The left plots are for $J=\frac{1}{2}$ and the right
    ones for $J=\frac{3}{2}$. For the two largest values of
    $\Lambda$, some resonant states from less attractive ${\rm
      SU(6)}_{\rm lsf}\times$HQSS multiplets are also visible for higher masses. This figure is taken from \cite{Nieves:2017jjx}.}
  \label{fig:lambdas}
\end{figure*}

Next, we aim at assessing the dependence of our results on the cutoff. Therefore, we examine higher and lower values of the cutoff. In Fig.~\ref{fig:lambdas}, we show the  pole positions for $\Lambda= 1090$ MeV (Table~\ref{tab:table3}) and two additional cutoffs, approximately 100 MeV apart.  We find that for cutoffs below 800 MeV, all states generated dynamically become heavier and wider than the experimental ones, whereas for bigger values than 1300--1350 MeV, our states appear well below 3 GeV.  Coming back to Fig.~\ref{fig:lambdas}, we find that  some (probably at least three) of the states observed by
LHCb~\cite{Aaij:2017nav} can be identified with three of our $\Omega_c^{(*)}$. In order to make the experimental identification possible,  a
significant coupling to the $\Xi_c \bar K$ channel has to be obtained,
often via $\Xi_c^* \bar K$ and $\Xi_c \bar K^*$ allowing for the $d-$wave transitions.

The molecular nature of the five $\Omega_c^{(*)}$ narrow states has been recently analyzed in
Refs.~\cite{Montana:2017kjw,Debastiani:2017ewu}, as previously mentioned, as well as the
observed broad  3188 MeV in Ref.~\cite{Wang:2017smo}. In Ref.~\cite{Montana:2017kjw} two $J=1/2$ baryon-meson molecular states could be
identified with the experimental $\Omega_c(3050)$ and
$\Omega_c(3090)$. These two states have been
reproduced in the $J=1/2$ sector in Ref.~\cite{Debastiani:2017ewu}, whereas  a $J=3/2$ molecular state has been also identified 
with the experimental $\Omega_c(3119)$, as the authors incorporate baryon $3/2^+$-pseudoscalar meson states into the hidden gauge approach. 
Compared to these works, our model for $\Lambda=1090$ identifies $J=1/2^-$ $\Omega_c(3000)$, $\Omega_c(3119/3090)$ and $J=3/2^-$ $\Omega_c(3050)$. This due to the fact that we use a different RS as well as  different interaction matrices, in particular for the channels involving $D$, $D^*$ and light vector mesons.

With regards to the broad structure observed by the LHCb Collaboration around 3188 MeV, the authors of Ref.~\cite{Wang:2017smo} have indicated that it could be the superposition of two $D \Xi$ bound states. In our case,  it is difficult to reach any identification with the experimental result, since most likely our model would also have to consider states from less attractive ${\rm SU(6)}_{\rm lsf}\times$HQSS multiplets \cite{Romanets:2012hm}. Also, a candidate of a loosely bound molecular state of mass around 3140 MeV is  predicted in Ref.~\cite{Chen:2017xat}, whereas in our case we cannot identify it with any of ours. We actually cannot associate our states to those of Ref.~\cite{Chen:2017xat}, as the authors do not consider $\Xi^{(*)}D^{(*)}$ channels.

\section{Conclusions}

In view of the recent LHCb experimental results~\cite{Aaij:2017nav}, we have revisited the $\Omega_c^{(*)}$ states by reviewing the RS used in the unitarized coupled-channel model
of Ref.~\cite{Romanets:2012hm} . In this previous work, five odd-parity
$\Omega_c, \Omega^*_c$ states were dynamically generated with masses below 3 GeV, thus making  the identification with any of the LHCb resonances difficult.

The predicted masses can be moved to higher energies by implementing a different RS. On the one hand,  the common energy-scale used in \cite{Romanets:2012hm} to perform the subtractions is modified allowing for moderate variations.  On the other hand,  a common UV cutoff is used to render finite the UV divergent loop functions in all channels. 

We  conclude that probably at least three of the states observed by LHCb~\cite{Aaij:2017nav} have  $J=1/2^-$ and
$J=3/2^-$. Indeed, those associated to the poles {\bf b} with $J=3/2$ and
{\bf c} with $J=1/2$ in Table~\ref{tab:table3} for $\Lambda=1090$ MeV would belong to the same  ${\rm SU(6)}_{\rm lsf}$ $\times$ HQSS multiplets as the  $\Lambda_c(2595)$ and
$\Lambda_c(2625)$ states, as well as $\Lambda_b(5912)$ and $\Lambda_b(5920)$ resonances.

\acknowledgments
L.T. acknowledges support from the Heisenberg Programme of DFG under the Project Nr. 383452331, the THOR COST Action CA15213 and the DFG through the grant CRC-TR 211. R. P. Pavao wishes to thank
the Generalitat Valenciana in the program Santiago Grisolia. This
research is supported by the Spanish Ministerio de Econom\'ia y
Competitividad and the European Regional Development Fund, under
contracts FIS2014-51948-C2-1-P, FIS2017-84038-C2-1-P, FPA2016-81114-P and SEV-2014-0398 and by Generalitat Valenciana under
contract PROMETEOII/2014/0068.

\bibliographystyle{JHEP}
          \bibliography{biblio}

\providecommand{\href}[2]{#2}\begingroup\raggedright\begin{thebibliography}{10}

\bibitem{Aaij:2017nav}
{\scshape LHCb} collaboration, R.~Aaij et~al., \emph{{Observation of five new
  narrow $\Omega_c^0$ states decaying to $\Xi_c^+ K^-$}},
  \href{https://doi.org/10.1103/PhysRevLett.118.182001}{\emph{Phys. Rev. Lett.}
  {\bfseries 118} (2017) 182001}
  [\href{https://arxiv.org/abs/1703.04639}{{\ttfamily 1703.04639}}].

\bibitem{Yelton:2017qxg}
{\scshape Belle} collaboration, J.~Yelton et~al., \emph{{Observation of Excited
  $\Omega_c$ Charmed Baryons in $e^+e^-$ Collisions}},
  \href{https://doi.org/10.1103/PhysRevD.97.051102}{\emph{Phys. Rev.}
  {\bfseries D97} (2018) 051102}
  [\href{https://arxiv.org/abs/1711.07927}{{\ttfamily 1711.07927}}].

\bibitem{JimenezTejero:2009vq}
C.~E. Jimenez-Tejero, A.~Ramos and I.~Vidana, \emph{{Dynamically generated open
  charmed baryons beyond the zero range approximation}},
  \href{https://doi.org/10.1103/PhysRevC.80.055206}{\emph{Phys. Rev.}
  {\bfseries C80} (2009) 055206}
  [\href{https://arxiv.org/abs/0907.5316}{{\ttfamily 0907.5316}}].

\bibitem{Hofmann:2005sw}
J.~Hofmann and M.~F.~M. Lutz, \emph{{Coupled-channel study of crypto-exotic
  baryons with charm}},
  \href{https://doi.org/10.1016/j.nuclphysa.2005.08.022}{\emph{Nucl. Phys.}
  {\bfseries A763} (2005) 90}
  [\href{https://arxiv.org/abs/hep-ph/0507071}{{\ttfamily hep-ph/0507071}}].

\bibitem{GarciaRecio:2008dp}
C.~Garcia-Recio, V.~K. Magas, T.~Mizutani, J.~Nieves, A.~Ramos, L.~L. Salcedo
  et~al., \emph{{The s-wave charmed baryon resonances from a coupled-channel
  approach with heavy quark symmetry}},
  \href{https://doi.org/10.1103/PhysRevD.79.054004}{\emph{Phys. Rev.}
  {\bfseries D79} (2009) 054004}
  [\href{https://arxiv.org/abs/0807.2969}{{\ttfamily 0807.2969}}].

\bibitem{Romanets:2012hm}
O.~Romanets, L.~Tolos, C.~Garcia-Recio, J.~Nieves, L.~L. Salcedo and R.~G.~E.
  Timmermans, \emph{{Charmed and strange baryon resonances with heavy-quark
  spin symmetry}},
  \href{https://doi.org/10.1103/PhysRevD.85.114032}{\emph{Phys. Rev.}
  {\bfseries D85} (2012) 114032}
  [\href{https://arxiv.org/abs/1202.2239}{{\ttfamily 1202.2239}}].

\bibitem{Montana:2017kjw}
G.~Montaña, A.~Feijoo and À.~Ramos, \emph{{A meson-baryon molecular
  interpretation for some $\Omega_{c}$ excited states}},
  \href{https://doi.org/10.1140/epja/i2018-12498-1}{\emph{Eur. Phys. J.}
  {\bfseries A54} (2018) 64}
  [\href{https://arxiv.org/abs/1709.08737}{{\ttfamily 1709.08737}}].

\bibitem{Debastiani:2017ewu}
V.~R. Debastiani, J.~M. Dias, W.~H. Liang and E.~Oset, \emph{{Molecular
  $\Omega_c$ states generated from coupled meson-baryon channels}},
  \href{https://doi.org/10.1103/PhysRevD.97.094035}{\emph{Phys. Rev.}
  {\bfseries D97} (2018) 094035}
  [\href{https://arxiv.org/abs/1710.04231}{{\ttfamily 1710.04231}}].

\bibitem{Gamermann:2010zz}
D.~Gamermann, C.~Garcia-Recio, J.~Nieves, L.~L. Salcedo and L.~Tolos,
  \emph{{Exotic dynamically generated baryons with negative charm quantum
  number}}, \href{https://doi.org/10.1103/PhysRevD.81.094016}{\emph{Phys. Rev.}
  {\bfseries D81} (2010) 094016}
  [\href{https://arxiv.org/abs/1002.2763}{{\ttfamily 1002.2763}}].

\bibitem{GarciaRecio:2012db}
C.~Garcia-Recio, J.~Nieves, O.~Romanets, L.~L. Salcedo and L.~Tolos, \emph{{Odd
  parity bottom-flavored baryon resonances}},
  \href{https://doi.org/10.1103/PhysRevD.87.034032}{\emph{Phys. Rev.}
  {\bfseries D87} (2013) 034032}
  [\href{https://arxiv.org/abs/1210.4755}{{\ttfamily 1210.4755}}].

\bibitem{Garcia-Recio:2013gaa}
C.~Garcia-Recio, J.~Nieves, O.~Romanets, L.~L. Salcedo and L.~Tolos,
  \emph{{Hidden charm N and Delta resonances with heavy-quark symmetry}},
  \href{https://doi.org/10.1103/PhysRevD.87.074034}{\emph{Phys. Rev.}
  {\bfseries D87} (2013) 074034}
  [\href{https://arxiv.org/abs/1302.6938}{{\ttfamily 1302.6938}}].

\bibitem{Tolos:2013gta}
L.~Tolos, \emph{{Charming mesons with baryons and nuclei}},
  \href{https://doi.org/10.1142/S0218301313300270}{\emph{Int. J. Mod. Phys.}
  {\bfseries E22} (2013) 1330027}
  [\href{https://arxiv.org/abs/1309.7305}{{\ttfamily 1309.7305}}].

\bibitem{Garcia-Recio:2015jsa}
C.~Garcia-Recio, C.~Hidalgo-Duque, J.~Nieves, L.~L. Salcedo and L.~Tolos,
  \emph{{Compositeness of the strange, charm, and beauty odd parity $\Lambda$
  states}}, \href{https://doi.org/10.1103/PhysRevD.92.034011}{\emph{Phys. Rev.}
  {\bfseries D92} (2015) 034011}
  [\href{https://arxiv.org/abs/1506.04235}{{\ttfamily 1506.04235}}].

\bibitem{Aaij:2012da}
{\scshape LHCb} collaboration, R.~Aaij et~al., \emph{{Observation of excited
  $\Lambda_b^0$ baryons}},
  \href{https://doi.org/10.1103/PhysRevLett.109.172003}{\emph{Phys. Rev. Lett.}
  {\bfseries 109} (2012) 172003}
  [\href{https://arxiv.org/abs/1205.3452}{{\ttfamily 1205.3452}}].

\bibitem{Nieves:2017jjx}
J.~Nieves, R.~Pavao and L.~Tolos, \emph{{$\Omega _c$ excited states within a
  $\mathrm{SU(6)}_{\mathrm{lsf}}\times $ HQSS model}},
  \href{https://doi.org/10.1140/epjc/s10052-018-5597-3}{\emph{Eur. Phys. J.}
  {\bfseries C78} (2018) 114}
  [\href{https://arxiv.org/abs/1712.00327}{{\ttfamily 1712.00327}}].

\bibitem{Nieves:2001wt}
J.~Nieves and E.~Ruiz~Arriola, \emph{{The S(11) - N(1535) and - N(1650)
  resonances in meson baryon unitarized coupled channel chiral perturbation
  theory}}, \href{https://doi.org/10.1103/PhysRevD.64.116008}{\emph{Phys. Rev.}
  {\bfseries D64} (2001) 116008}
  [\href{https://arxiv.org/abs/hep-ph/0104307}{{\ttfamily hep-ph/0104307}}].

\bibitem{Wang:2017smo}
C.~Wang, L.-L. Liu, X.-W. Kang, X.-H. Guo and R.-W. Wang, \emph{{Possible
  open-charmed pentaquark molecule $\Omega_c(3188)$ - the $D \Xi$ bound state -
  in the Bethe-Salpeter formalism}},
  \href{https://doi.org/10.1140/epjc/s10052-018-5874-1}{\emph{Eur. Phys. J.}
  {\bfseries C78} (2018) 407}
  [\href{https://arxiv.org/abs/1710.10850}{{\ttfamily 1710.10850}}].

\bibitem{Chen:2017xat}
R.~Chen, A.~Hosaka and X.~Liu, \emph{{Searching for possible $\Omega_c$-like
  molecular states from meson-baryon interaction}},
  \href{https://doi.org/10.1103/PhysRevD.97.036016}{\emph{Phys. Rev.}
  {\bfseries D97} (2018) 036016}
  [\href{https://arxiv.org/abs/1711.07650}{{\ttfamily 1711.07650}}].

\end{thebibliography}\endgroup
%\begin{thebibliography}{99}
%\bibitem{...}
%....
%\end{thebibliography}

\end{document}